# Bonding mechanisms in cold spray deposition of gas atomised and solution heat-treated Al 6061 powder by EBSD

Alexandre Sabard, Tanvir Hussain*

Faculty of Engineering, University of Nottingham, Nottingham, NG7 2RD, UK
*Corresponding author: tanvir.hussain@nottingham.ac.uk, +44 115 951 3795

**Abstract**

The heat-treatment of a number of gas atomised aluminium alloys prior to cold spraying recently showed that the resultant microstructural modification was accompanied by an improvement in deposition; however, the relationship between the microstructural homogenisation occurring after recrystallisation and the increase in deposition efficiency and particle-particle bonding had not been investigated. In this study, Al 6061 gas atomised feedstock powder, before and after solution heat-treatment, was cold sprayed and these materials were characterised using electron backscatter diffraction. The solution heat-treated Al 6061 powder showed large stress-free grains as opposed to the as-atomised feedstock powder which exhibited smaller grains with the presence of dislocations. The coating produced from as-received powder exhibited a homogeneous distribution of misorientation and lattice defects throughout the particles, whereas the coating produced from solution heat-treated powder showed an accumulation of dislocations in the interfacial zones. For the first time, a mechanism was proposed where this phenomenon was attributed to localised deformation of the heat-treated particle's exterior due to the dissolution of precipitates in the intermetallic network. The accumulation of dislocations in the interfacial area enhanced the rotation of subgrains to accommodate the plastic deformation, thus improving the particle-particle bonding.

**Introduction**

Cold gas dynamic spray is a well-established materials processing technique that is based on the kinetic energy of incident particles to deposit coatings. Cold spray uses metallic powders accelerated by a supersonic gas through a convergent-diverging nozzle (also called De Laval) and the particles impact onto a surface, and the combined plastic deformation of the feedstock material and the substrate forms a coating [1–3]. Cold spray is a solid-state process, where every sprayed material is kept below its melting temperature and therefore any deleterious effect of the temperature during deposition is avoided [4,5].

Gas atomised powders is the most common type of powder used in the recent cold spray applications and its microstructure was extensively examined recently in relation to cold spray. Due to the nature of the gas atomisation process and being closely related to the high cooling rates ($10^{-4}$ to $10^{-8}$ $K.s^{-1}$[6]) reached by the melt during solidification, the powders typically exhibit a cellular/dendritic microstructure [7,8]. Attributed to several mechanisms [9], the solute atoms are rejected from the matrix during the dendrite growth [10] forming a brittle intermetallic cellular network in a variety of materials, from Fe-based alloys [11] to Al alloys [12, 13]. The presence of an inconsistent intermetallic network can affect the formability of the powder and has been shown to affect any post-deposition aging heat treatment [14]. To solve this issue, the heat treatment of the powder prior to spraying was recently considered and performed on several gas atomised aluminium alloys, such as Al-Zn-Mg, Al-Mg-Si and Al-Cu alloys [15–17]. Based on the diffusion of elements throughout the matrix during a solution heat treatment cycle above the solvus curve, the solute atoms segregation was reduced, and the non-uniformly distributed alloying elements were mostly dissolved in the aluminium solid solution. The homogenisation of the microstructure examined was accompanied by an increase in deposition efficiency and a stronger bonding between particles as well as with the substrate [15,16].

Over the past decades, several models were used to describe the cold spray bonding mechanisms. Predominantly depending on the ductility of the substrate and feedstock material, the successful deposition of a powder relies on various parameters [18–20]. A "critical velocity" was thus defined as the speed the particles are required to reach in order to adhere to the surface on which they are sprayed, defined as a function of the sprayed material characteristics [2]. Above this critical velocity, the materials undergo a high strain rate deformation in the impact area where the thermal softening effect associated with the deformation lead to adiabatic shear instability in this region. Subsequently, an interfacial jet is formed and removes the oxide layer from the particle surface, facilitating an intimate bond between two particles or the particles and the substrate [21].

Moreover, it is generally accepted that the accumulation of dislocations during severe plastic deformation is responsible for various grain structure formations observed in cold sprayed deposits: from elongated grains (200 nm to 1 µm) to ultrafine grains (50 to 200 nm) [5,9,19] in the interfacial regions. The dislocation piling into arrays of dislocation walls during the high strain rate plastic deformation and the rotation of the lattice along the shear leads to the formation of elongated grains [23]. Subsequently, these elongated grains are subdivided into smaller subgrains due to further accumulation of dislocations. In order to accommodate the intensive plastic deformation in the impact area, the newly-formed subgrain rotate and form equiaxed ultrafine grains at the particle/particle interface. More recently, the interaction of powder constituted of a second-phase was studied and the interaction of the second phase with the dislocations during deposition was determined to be linked with the formation of ultrafine grains [24]. Although nano-sized grains were also observed in Ti cold spray coatings in the particle interior, the formation of ultrafine grains in the particle/particle interface was ascertained in many cases [5] and was showed to contribute to the metallurgical bonding [25].

The grain structure of a cold spray coating is an indicator of its deposition mechanism, and Electron Backscatter Diffraction (EBSD) was proven to be a useful tool in order to observe the microstructure of the deposits [14,20,22]. Although the indexing of grains at the particle/particle interface revealed itself challenging due to the heavy lattice deformation occurring in this area, EBSD scans of cold spray coatings provide a various range of data which can be interpreted. Low-angle and High-angle grain boundaries distribution [22,23] as well as local average misorientation [5] are crucial to determine the strain localisation or the dislocation density.

Although the microstructure of cold spray deposits was widely studied, the effect of a heat-treatment on the powder on its deposition on a substrate is yet to be observed on a microstructural level. The improvement of deposition was already proved in previous studies without further explanation. The grain structure alteration of the particle of powder, as well as a potential stress release during heat treatment are two factors which matter during cold spray deposition due to the solid-state nature of the process.

In this study, a solution heat treatment was performed on an Al 6061 powder and its effect was analysed on a microstructural level using SEM and EDS. The particles grain structure and its size distribution before and after heat treatment was also characterised using EBSD and the resulting coatings were also analysed to understand the various deposition mechanisms occurring cold spray deposition. The misorientation and the grain boundaries density was compared in order to evaluate the effect of the heat-treatment on the particle plastic deformation.

**Experimental methods**

1. **Materials**

A spherical gas atomised Al 6061 powder (LPW, Runcorn, UK) was the feedstock material of this study. The chemical composition of the powder was 1.00 wt.% Mg, 0.65 wt.% Si, 0.26 wt.% Cr, 0.22 wt.% Cu, 0.1 wt.% Mn, 0.07 wt.% Fe and 0.01 wt.% Zn, the remainder being Al. The substrates used for the cold spray experiments were 3 mm thick 6061-T6 aluminium alloy plates (30x100 mm). The composition of the substrate is 0.99 wt. % Mg, 0.66 wt. %Si, 0.16 wt.% Cr, 0.31 wt.% Cu, 0.08% Mn, 0.25% Fe and 0.01 wt.% Zn, the remainder being Al. T6 corresponds to the heat treatment applied to the material, which is in that case a solution heat treatment followed by an artificial aging. The particle size analysis was performed using laser diffractometry (Laser Mastersizer 3000, Malvern Instruments, Malvern, UK) where the metallic powder was dispersed in air.

2. **Solution Heat treatment and quenching of the powder feedstock**

The solution heat treatment of the metallic powder was performed under controlled conditions due to its explosivity. In the presence of an ignition source and oxygen, various metal powders can blow up if they are at a minimum temperature called ignition temperature. In the solution heat treatment cycle used to homogenise the powder, Al 6061 was kept at 530°C for 4 hours. The temperature was chosen following the standard T6 (solution heat treatment + artificial aging) conditions [28]. In order to avoid any fire hazard, 120 g of the feedstock powder was sealed in a quartz tube prior to their heat treatment in order to prevent any exposure to oxygen during the following quench. A 10-mPa vacuum was created in the vial using a diffusion pump before the sealing was performed using an oxy-propane flame, closing the top of the vial. The dimensions of the tube after sealing were as follows: outer diameter 14 mm, length ~100 mm and 2 mm wall thickness. The quartz tube was then inserted in the oven and kept at 530°C for 4 hours. Subsequently, the tube containing the powder was quenched into water for 2 minutes until it reached room temperature. In this case, the solid solution obtained during the heating cycle is quenched to be maintained in this state and to limit the diffusion during cooling. Once at room temperature, the vial was opened, and powder collected for further analysis and experiments. The powder heat treatment and its cold spray deposition were performed the same day in order to prevent any aging from occurring.

3. **Cold spray deposition**

The coating deposition was performed using a bespoke high pressure cold spray system at the University of Nottingham. The rig setup was described in more detail in a previous publication [29]. The coatings were deposited onto Al 6061 substrates at He gas pressure of 2.9 MPa at room temperature. A commercial high-pressure powder feeder (Praxair 1264 HP, Indianapolis, IN, USA) was used at a pressure 0.1 MPa higher than the main pressure. A hardened stainless nozzle was used for the experiments, designed with an area expansion ratio of 8 and a divergent length of 150 mm. The nozzle was held stationary while the substrates were maintained on a x-y table. The substrate was located at 20 mm from the nozzle exit and 4 passes were sprayed onto the substrate by using a transverse speed of 60 mm/s in order to obtain the required thickness for an analysis of the coating deposition. An overlap of 2 mm was set to deposit a coating homogeneous in thickness.

4. **Material Characterisation**

The morphology of the powder was observed using a scanning electron microscopy (JEOL JSM 6490LV, JEOL Ltd., Japan) operating at 10 kV. The cross-section of the feedstock powder and the deposits were evaluated using a Field-emission gun SEM (7100F, JEOL Ltd., Japan) at an accelerating voltage of 15 kV. The cold sprayed samples were sectioned using a precision saw at a cutting speed of 0.005 mm/s to avoid any damage on the coating during the operation. The powders and cross-sectioned cold spray deposits were mounted in epoxy resin and subsequently ground with P1200 silicon carbide paper, whereas cold spray deposits were ground using P240, P400, P800 and P1200 grinding paper. Once ground, samples were polished using 6 and 1 μm diamond paste. Final polishing was performed using a colloidal silica (0.06 μm) suspension. The microhardness analysis was performed using a MMT-7 Vickers Microhardness instrument (Buehler, IL, USA). Each sample underwent 8 measurements. A 10-gf load was applied for 10 s for mounted particles and coatings in cross-section.

## 5. Electron backscatter diffraction

EBSD was performed on a FEG SEM (7100F, JEOL Ltd., Japan) at an accelerating voltage of 15 kV. EBSD scans were carried out using AZtechHKL software (Oxford Instruments, UK). The data obtained were post-processed using the HKL Channel 5 Tango software (Oxford Instruments, UK). A noise reduction of a factor 3 was carried out, corresponding to the examination of each non-indexed pixel. If the non-indexed pixel has more than a specified number of indexed neighbours (in this case 6), it will be replaced by the most common neighbouring orientation [30,31]. High-Angle grain boundaries (>15°) and Low-angle boundaries (2-15°) are respectively represented on the Euler maps by black lines and white lines. Misorientation lines were also plotted to measure the orientation variation across particles. Point-to-point curves indicate the profiles of the orientation fluctuations between neighbouring points, whereas point-to-origin curves indicate the profiles of the orientation between all points and a reference point. Moreover, the local misorientation maps were plotted using AZtechHKL software. The average misorientation between each measurement and its 8 neighbours were calculated, excluding grain boundaries higher than 5° in angle. Typically used to localise strain variations, a colour gradient is used from blue corresponding to an absence of misorientation to a yellow/green colour indicating a high misorientation.

# Results

## 1. Solution heat treatment of the feedstock powder

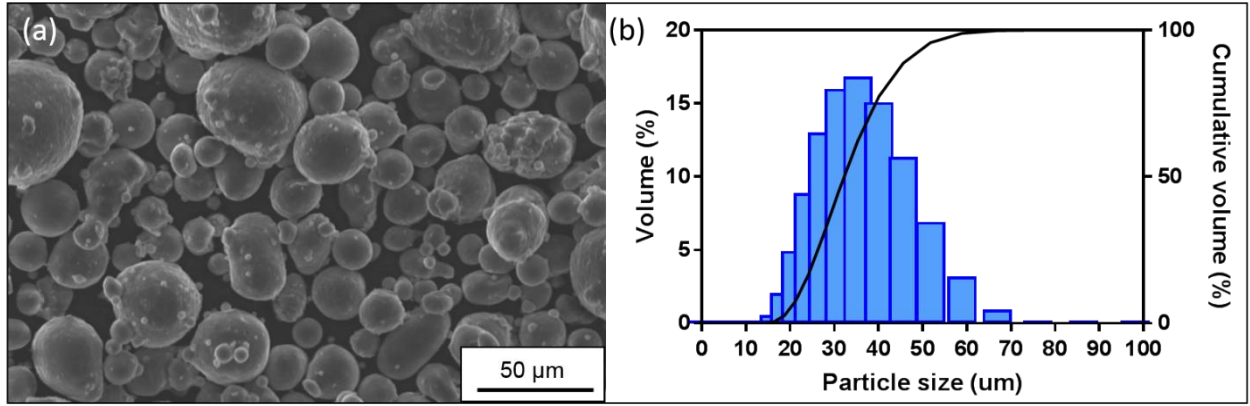

Fig. 1: (a) Topography of the feedstock Al 6061 powder showing a mostly spherical morphology with some satellite particles and (b) Particle size distribution measured by laser diffractometry showing the majority of particles between 30 and 40 micrometre diameters.

The feedstock powder exhibits a spherical shape, accompanied with some irregular particles; some much smaller particles can be also seen on the top of larger ones, called satellites. This phenomenon is attributed to the in-flight contact of molten droplets of different sizes. Small particles interact with larger ones under turbulent atomisation and welding occurs, leading to satelliting [32]. The particle size distribution is also displayed in Fig. 1(b). A gaussian distribution is observed with most of the particles between 30 and 40 µm of diameter. The $D_v10$, $D_v50$ and $D_v90$ of the Al 6061 powder batch was measured as 24.9, 36.5 and 53.0 µm respectively according to the particle size distribution results. These values correspond to the maximum particle diameter below which 10, 50 and 90% respectively of the sample volume exists.

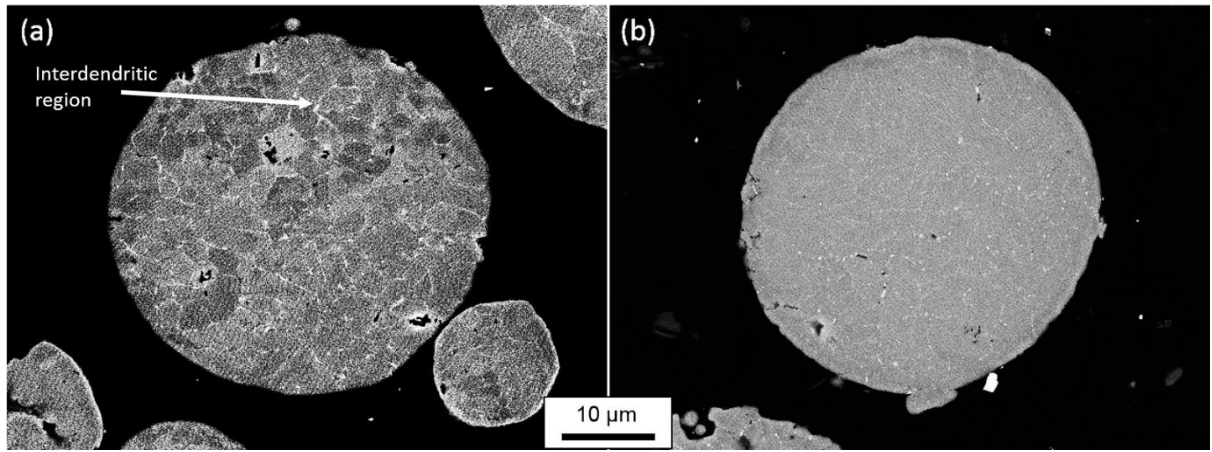

Fig. 2: BSE micrograph of the polished cross-section of (a) the feedstock Al 6061 powder and (b) the powder after solution heat treatment showing a homogenisation of the microstructure. The solute elements, namely Mg, Si, Cu, Fe and Cr have been redistributed in the matrix and the microstructure was retained by rapid quenching to avoid any diffusion during cooling.

The cross-section of the feedstock powder, in its feedstock state as well as in its heat-treated state is presented in Fig. 2. The feedstock powder exhibits a dendritic microstructure, depicted by a bright contrast observed on the BSE image in between the dendrites (Fig. 2 (a)). The brighter contrast reveals the potential presence of heavier elements in the interdendritic region (white arrow), as compared to the aluminium appearing darker in the matrix. This type of microstructure is typical of gas atomised metallic powders, and more generally rapidly solidified materials which tend to undergo solute segregation during solidification [33]. During the solidification, the relatively high cooling rate undergone by the droplet trigger its nucleation. The growth of α-Al from those nucleation points is also associated with a solute rejection, where the alloying elements are segregated as a solid solution on the sides of the dendrites [10], forming the microstructure observed in Fig. 2(a). The cross-section of the solution heat-treated particle (Fig. 2(b)) depicts a more homogeneous microstructure. The lack of contrast observed highlights that the solute atoms have been dissolved into solid solution leading to a partial homogenisation of the elemental distribution. A similar operation was performed on an AA7075 alloy in a previous study [15]; however, few precipitates can be observed suggesting that this homogenisation was not complete, possibly correlated to an insufficient cooling rate allowing enough diffusion for precipitates to form, or the heat-treatment was not sufficient to dissolve them into the aluminium matrix in the first place.

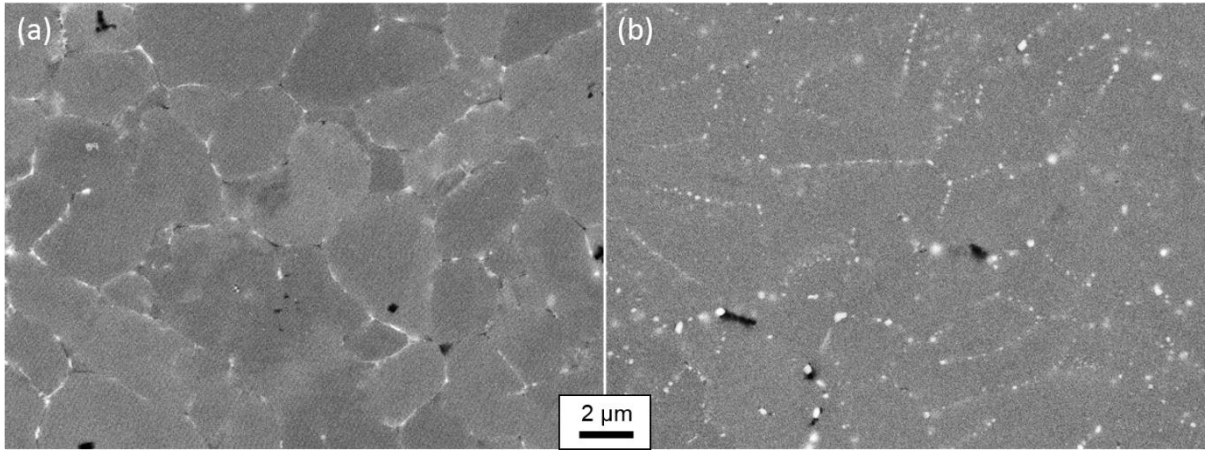

Fig. 3: SEM BSE micrographs at higher magnification showing partial homogenisation after solution heat treatment. The interdendritic bright regions of the feedstock powder (a) have been replaced by precipitates post-heat treatment (b).

Higher magnification micrographs reveal distinctive features on the powder cross-section (Fig. 3). The feedstock powder exhibits a continuous bright line (Fig. 3(a)) outside each dendritic cell corresponding to a solid solution of the solute atoms, in this case mainly Si, Mg, and Cu; however, the post-heat treatment microstructure depicted on Fig. 3 (b) shows discontinuous line made of extremely small precipitates.

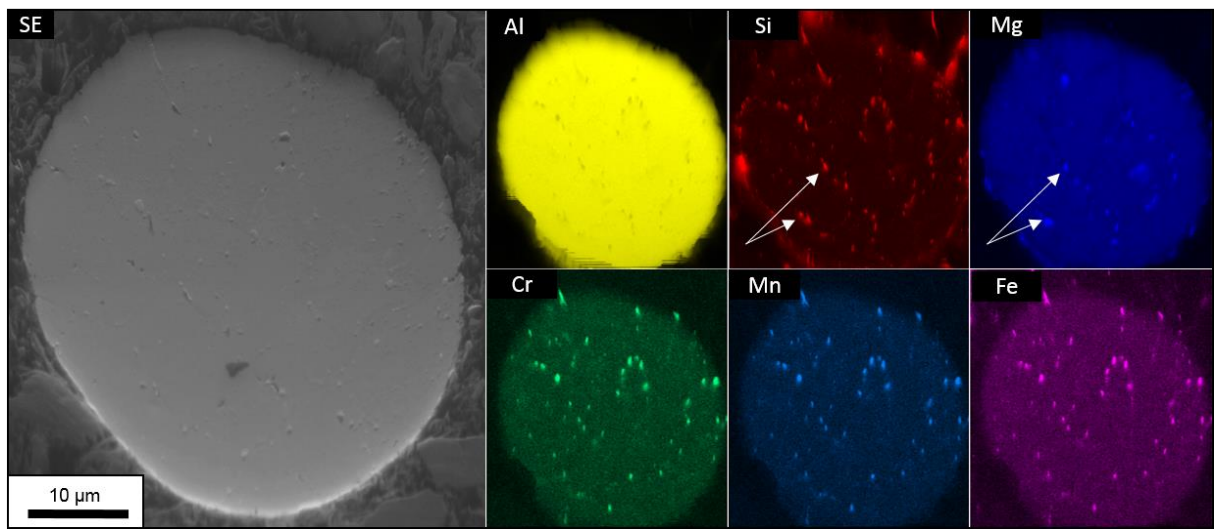

Fig. 4: Energy dispersive X-ray spectroscopy (EDS) of a solution heat-treated powder particle showing the elemental distribution. Precipitates composed of Si, Cr, Mn and Fe are distinguished in the microstructure, while white arrows also indicate the presence of Mg- and Si-rich only precipitates.

An elemental compositional analysis of a solution heat treated Al 6061 (Fig. 4) particle was performed to identify the precipitates in the microstructure. Cr, Mn, Fe and are observed to have the exact same distribution throughout the particle. Si is also observed in the same location which possibly suggests the presence of (Fe, Mn, Cr)$_3$SiAl$_{12}$ in the particle cross-section, precipitates previously reported in the homogenised 6000 series aluminium alloys [29,30]. The Mg distribution across the particle cross-section is, however, different from the previously mentioned element, the presence of Si is still noticed in the same areas. The presence of Mg$_2$Si is thus suggested, since similar precipitates were also reported in heat-treated 6000 series Al alloys [35].

## 2. Grain distribution

In order to get a more thorough understanding of the effect of the heat treatment on the powder particle microstructure, the grain distribution and orientation was observed using EBSD.

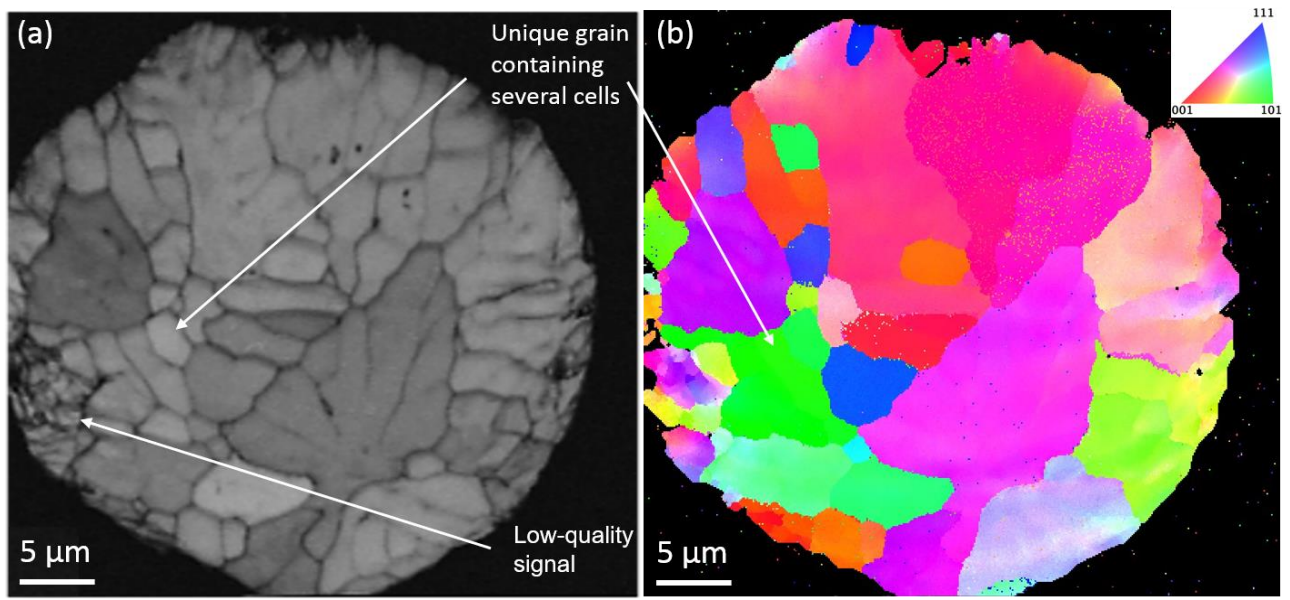

Fig. 5: EBSD characterisation of the Al 6061 powder. Band contrast (a) and Inverse Pole Figure in the Z direction (b) of a particle showing an irregular grain size distribution and shape. Large grains appear to be containing several dendritic cells. The band contrast image (a) present black lines at the particle exterior suggesting lattice irregularities.

The band contrast (BC) image from the EBSD scan of a feedstock particle of Al 6061 powder is presented in Fig. 5(a) which shows the quality of the diffracted signal. A bright colour corresponds to an undeformed material providing a sharp projection of the lattice whereas a darker colour corresponds to a lower quality of signal, generally attributed to the presence of a grain boundary or lattice defects. Surface preparation can also affect the band contrast, explaining the presence of dark lines on the particle edges due to polishing. Most of the dark lines appearing on the band contrast image of the feedstock powder correspond to grain boundaries, or interdendritic regions. Few randomly oriented black lines seem to be present at the centre-left of the particle, suggesting the potential presence of lattice irregularities. The difference in orientation between two grains is responsible for the variations in band contrast. The sharpness of the Kikuchi bands compared to the background will differ depending on the grain orientation itself. The Inverse Pole Figure (IPF) (Fig. 5(b)) attributes diverse colours to different grain orientation according to a specific direction, here being Z axis (see insert). Diverse sizes and shapes of grains can be distinguished, and some of the grains seem to contain several dendritic cells. The particle microstructure consists of elongated and irregular grains as well as smaller equiaxed ones. A leaf-like grain can also be observed in purple at the centre of the particle, a common feature in gas atomised powder, representing the growth of dendrites from a nucleation point during solidification.

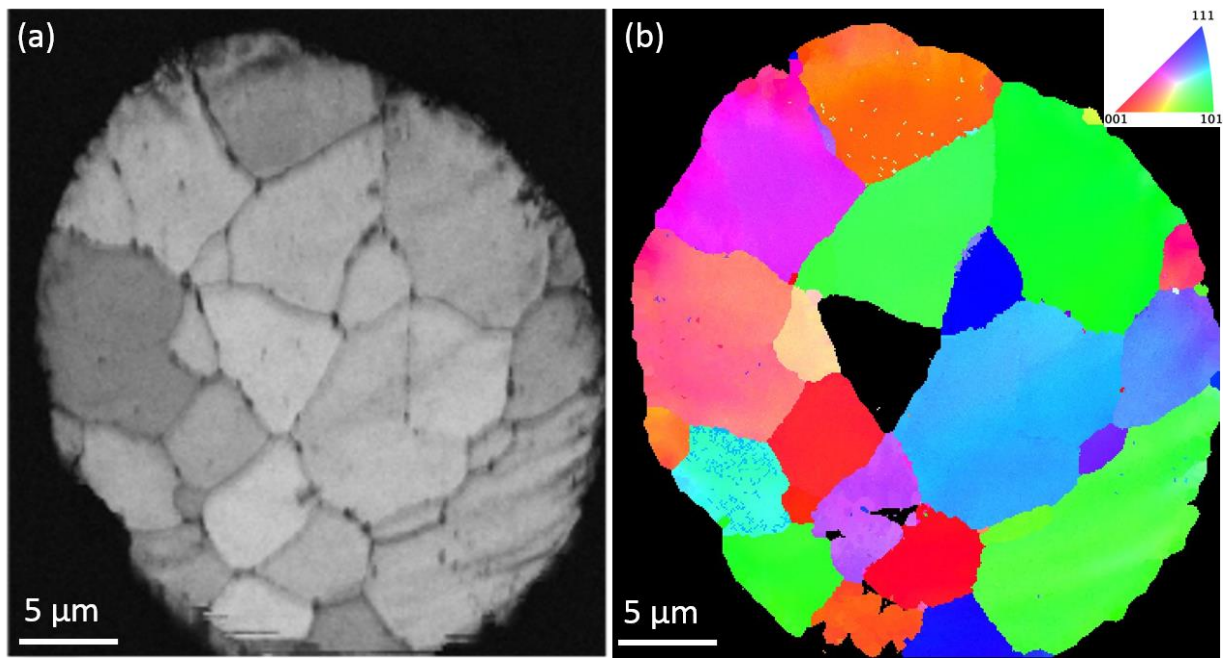

Fig. 6: EBSD characterisation of a solution heat treated Al 6061 particle. Band contrast (a) and Inverse Pole Figure in the Z direction (b) showing a good pattern quality as well as large equiaxed grains.

EBSD scan was also performed on the solution heat treated Al 6061 powder and the band contrast image (Fig. 6(a)) depicts a different structure in comparison with the feedstock powder (Fig. 5(a)). The particle exhibits a homogeneous distribution of large equiaxed grains. Slightly darker lines can be distinguished at the bottom right of the particle due to poor pattern quality, which can be attributed to polishing artefact. The inverse pole figure of this particle illustrates the various orientations of the grains.

A grain at the centre of the particle remain unindexed, although having a high band contrast. This was found to be attributed to a variation in lattice parameter of aluminium in the centre grain due to precipitates or solute atoms, which made the indexing impossible or reduced during the EBSD experiments.

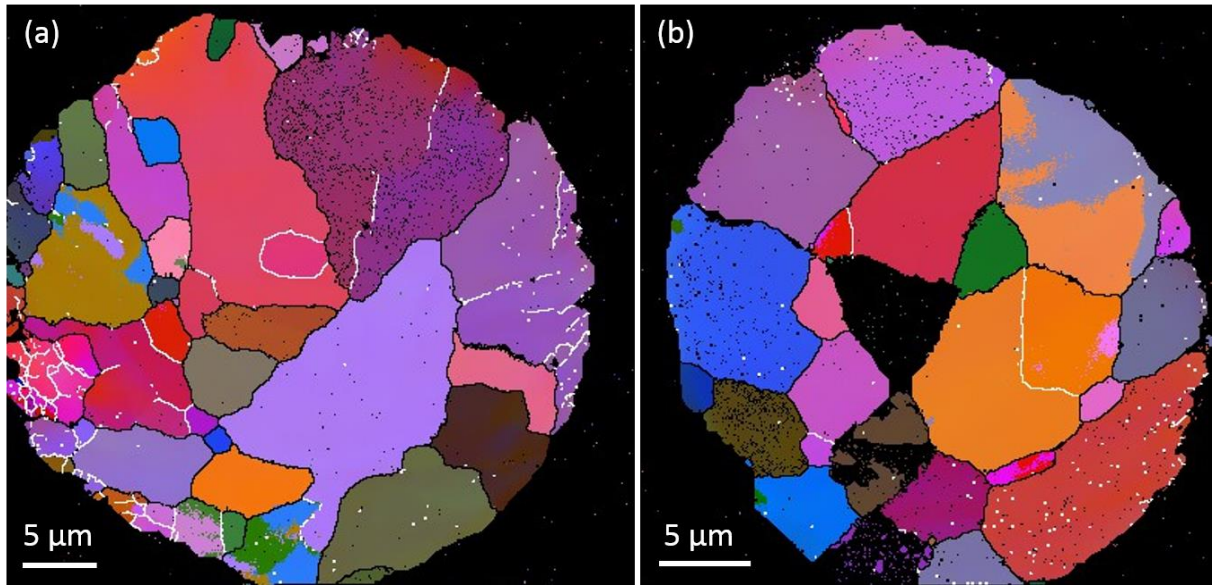

Fig. 7: Euler maps showing the LAGB (between 1 and 15°, white lines) and HAGB (>15°, black lines) distribution in the Al 6061 powder before (a) and after heat-treatment (b). Substantial number of LAGB are observed in the feedstock powder suggesting presence of dislocation.

Euler scans of the powder before and after heat treatment are presented Fig. 7. In attributing a colour to each Euler angle, this figure represents the grain orientation relative to a reference coordinate system in combining the value of the three angles to a single RGB colour [36]. The Euler scans were here chosen to be presented as they reveal the orientation of the grains on a 3-axis system, more detailed than the Inverse Pole Figure. On those pictures, Low-Angle Grain Boundaries (LAGB) and High-Angle Grain Boundaries (HAGB) are added to the scans in order to understand further the effect of the heat treatment. The feedstock Al 6061 present a wide number of LAGB at its extremities and especially in the bottom-left part. By contrast, the heat-treated powder particle is mainly constituted of HAGB, presenting only one major LAGB in the centre of the particle. The distinction between LAGB and HAGB is crucial for understanding EBSD data, since LAGB are generally described as an array of dislocations [37] and thus are representative of an accumulation of lattice defects in a material. On the other hand, HAGB are considered independent of the misorientation and are a characteristic of newly formed grains, such as recrystallised grains [38].

## 3. Cold spray deposition

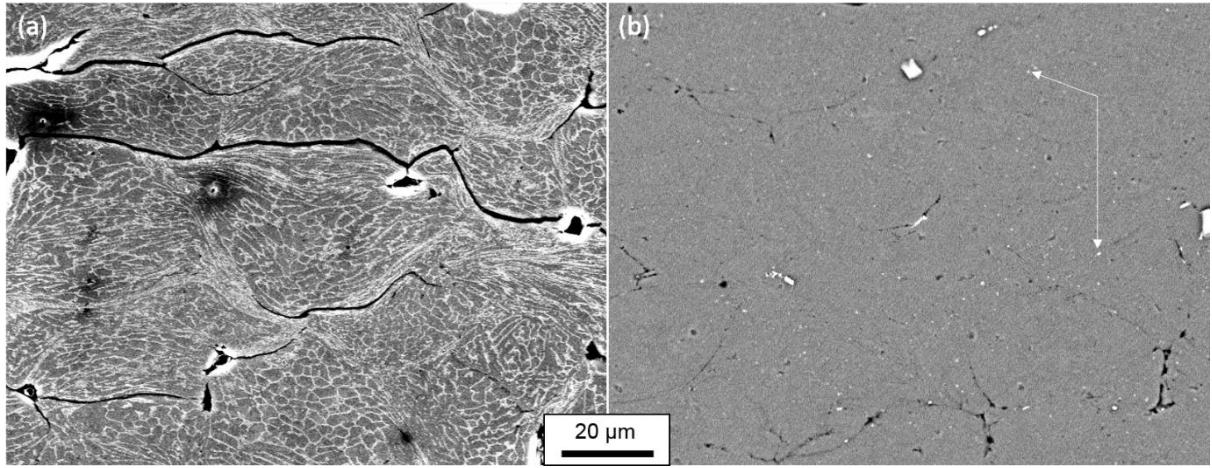

Fig. 8: BSE micrographs of cold spray deposits using (a) the feedstock Al 6061 powder and (b) the heat-treated powder. The first deposit (a) exhibits a dendritic microstructure as well as large gaps between particles; however, the second deposit shows a homogenised microstructure accompanied with a closer particle/particle bond.

The coating sprayed using the feedstock powder (Fig. 8(a)) exhibits the dendritic microstructure observed previously on the powder cross-section. The deformation of the incident particles is well-illustrated by the bright interdendritic regions appearing compacted in the interfacial regions. Gaps can be observed in between the splats, indicator of poor inter-particle bonding. Several triple junction voids can also be observed, indication of a lack of deformation in the process.

The coating deposited using the solution heat treated powder (Fig. 8(b)) presents a homogeneous microstructure, already observed in the heat-treated particles. The nanometric precipitates seen using EDS and SEM are also present in the coating microstructure. An intimate bonding is observed between the majority of the particles, indication of a better cohesion in the coating.

Overall, the feedstock and the solution heat treated powder exhibited two different behaviours. A better deposition was observed in the case of the heat-treated powder, revealing an intimate bonding with limited porosity; on the other side, a much higher porosity is noticed in the coating produced from feedstock powder. The gaps surrounding most of the particles is a clear indicator of a particle impinging on a previously deposited material which is deforming but not forming a sufficiently strong bond to create a pore free coating.

The microhardness of both powders prior to the process were measured as well as a value in the centre or the resulting coatings. The feedstock Al 6061 powder was measured at 105.3 ± 3.2 HV, whereas the hardness of the particles post-heat treatment was evaluated at 82.5 ± 5.6, 24 hours after spraying. The coatings hardness was also measured 24 hours after spraying, where the coating deposited using feedstock Al 6061 powder was measured at 135 ± 6.2 and the one sprayed using the pre-processed powder resulted in a coating at 123.5 ± 3.6 HV. The hardness values highlight the work-hardening occurring during deposition.

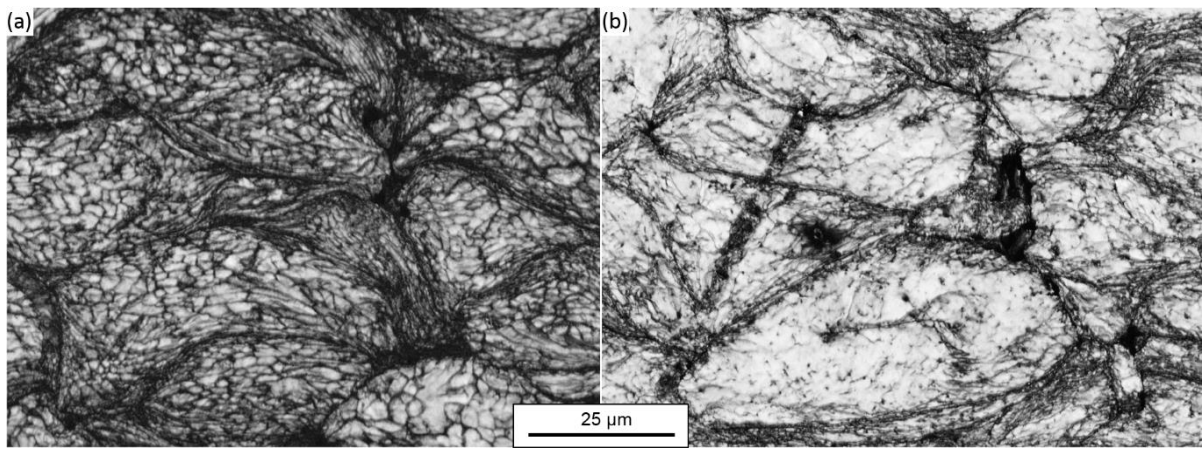

Fig. 9: Band contrast images of cold spray deposits using (a) feedstock powder and (b) solution heat treated powder. A better pattern quality is observed in the second case (b) suggesting a lower amount of lattice defects compared to the coating (b) which exhibits a darker contrast due to a high number of grain boundaries.

EBSD scans were performed on both coatings in order to observe the grain distribution and the lattice deformation in the deposits. The Band contrast (Fig. 9) is an important indicator in cold spray deposition since it will be affected by the presence of dislocations as well as the grain deformation. An area of the same size was chosen for EBSD scans in both cases, located in the centre of the coatings (approximately 100 µm below the top surface of the coating and 100 µm above the coating-substrate interface). In the case of the coating deposited using the feedstock powder, each splat is easily distinguished, thanks to the lack of indexing at the particle/ particle boundaries appearing in black due to the low quality of the signal. Moreover, the interdendritic regions already observed on the cross-section of the particle (Fig. 5(a)) also appear in black because of its composition. A small particle in the centre of the scanned area appears to be extremely deformed according to the deformation of the dendritic cells. On the band contrast image of the coating deposited using the heat-treated Al 6061 powder (Fig. 9(b)), a much brighter image is seen, representative of a better signal in the centre of the particles; however, a darker contrast is still observed at the particle-particle boundaries showing a high deformation in this area, limiting any indexing or grain orientation detection.

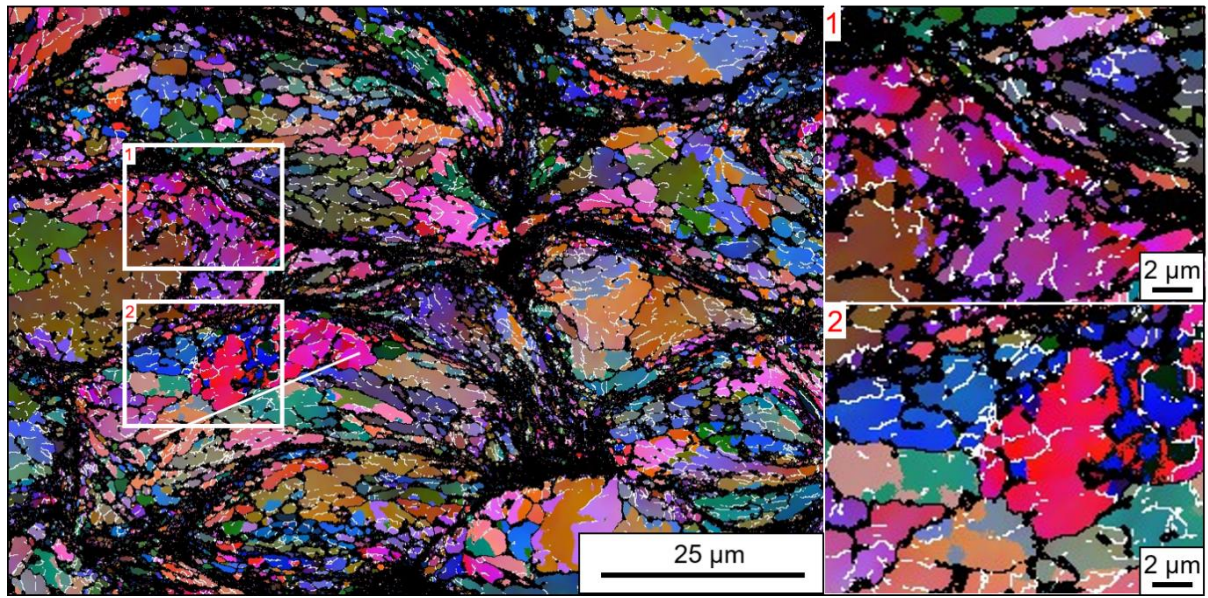

Fig. 10: Euler EBSD map of cold spray deposit using the feedstock Al 6061 powder showing a large distribution of LAGBs (between 1 and 15°, white lines) both in the interior and outside of the particles. The inserts are higher magnification images of the areas in the rectangles. The white line corresponds to the misorientation line (a) plotted Fig. 12 from left to right.

The Euler maps of the coatings deposited using the feedstock powder is displayed Fig. 10. The small grain size observed in the single particle EBSD scan (Fig. 5) can be also easily noticed on this coating cross-section. Moreover, the lattice deformation induced by the impact during deposition is visible through a change of colour inside an apparent grain. Distinct colours representing different orientation can be observed in several dendritic cells, showing the effect of the particle deformation and thus the lattice distortion during deposition. LAGBs (white lines) can be observed in the whole coating, widely distributed throughout each particle and most of the dendritic cells. The higher magnification pictures reveal that no clear difference exists between the particle interior and the particle-particle boundaries in terms of LAGB distribution.

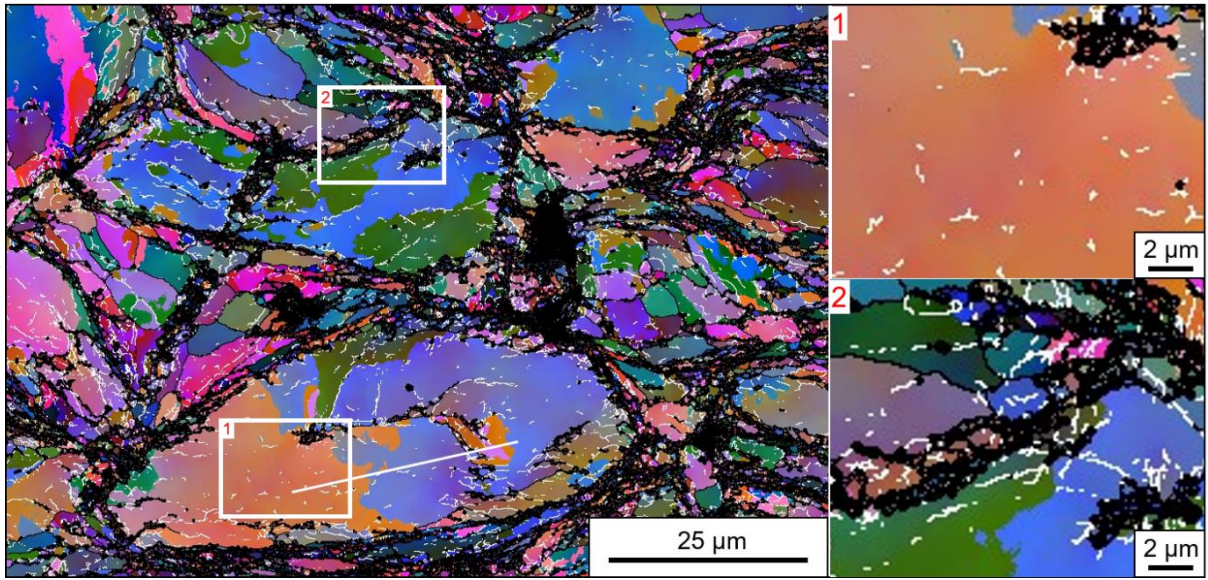

Fig. 11: Euler EBSD map of cold spray deposit using the solution heat treated Al 6061 powder showing LAGBs (between 1 and 15°, white lines) mostly located at the particle-particle boundaries. Insert 1 shows a particle interior containing a limited amount of LAGB whereas Insert 2 located at the particle-particle interface presents a wide distribution of LAGB suggesting an accumulation of lattice defects in this area. The white line indicates the misorientation line (b) plotted Fig. 12 from left to right.

On the other hand, the Euler map of the coating sprayed using the heat-treated powder in Fig. 11 shows a different behaviour: once again LAGBs can be observed throughout the coating but these tend to be located in the particle extremities. The inserts show a higher magnification image of a particle-particle boundary (insert 1) as well as a particle interior (insert 2), and LAGBs are agglomerated at the particle-particle interface whereas their presence is limited inside the particle. The presence of large grains provides a better illustration of the lattice deformation, the large particle observed at the bottom of the Euler figure exhibiting a colour change from orange to purple in what appears to be the same grain.

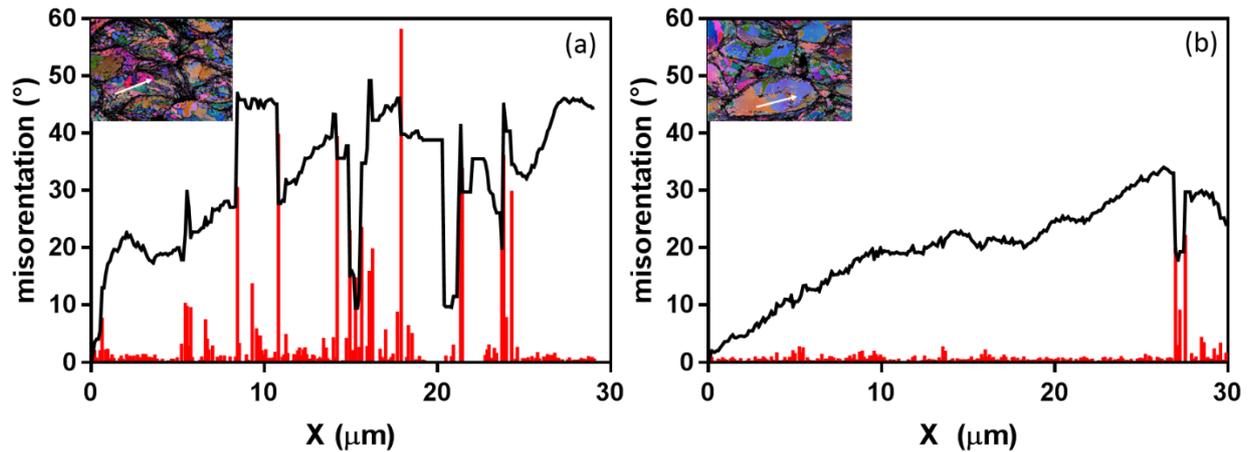

Fig. 12: Misorientation profiles of the particle interior in the cold spray deposited using (a) feedstock Al 6061 powder and (b) the solution-heat treated powder. Point to point (red bars) are plotted as well as point to origin (black line). The graphs (a) and (b) correspond respectively to the arrows indicated Fig. 10 and Fig. 11. Overall, the feedstock particle interior presents a large number of grains indicated by the HAGB as well as misorientation and LAGB across the whole particle. On the other hand, the heat-treated particle presents a limited misorientation suggesting a low lattice strain and dislocation density.

The misorientation lines taken in the splat interior (Fig. 12) provides a more accurate vision of the LAGB and HAGB distribution across a particle in the deposit. A large number of HAGB (>15°C) are measured across the feedstock powder particle, indicated clearly by the point to point red bars on the graph Fig. 12(a). In between each HAGB, a substantial amount of misorientation and LAGB is observed suggesting a heavy lattice strain and dislocation density in this area. On the other hand, the heat-treated particle (Fig. 12(b)) exhibits a low amount of misorientation, the point-to-origin black line displaying only a slow orientation gradient across the particle due to the lattice compression. It is also important to notice that this gradient is observed in both cases, similar to the colour gradient previously observed in the Euler maps (Fig. 10 and Fig. 11). Peaks are also observed in the red curves, corresponding to the presence of many non-indexed pixels in the plotted line.

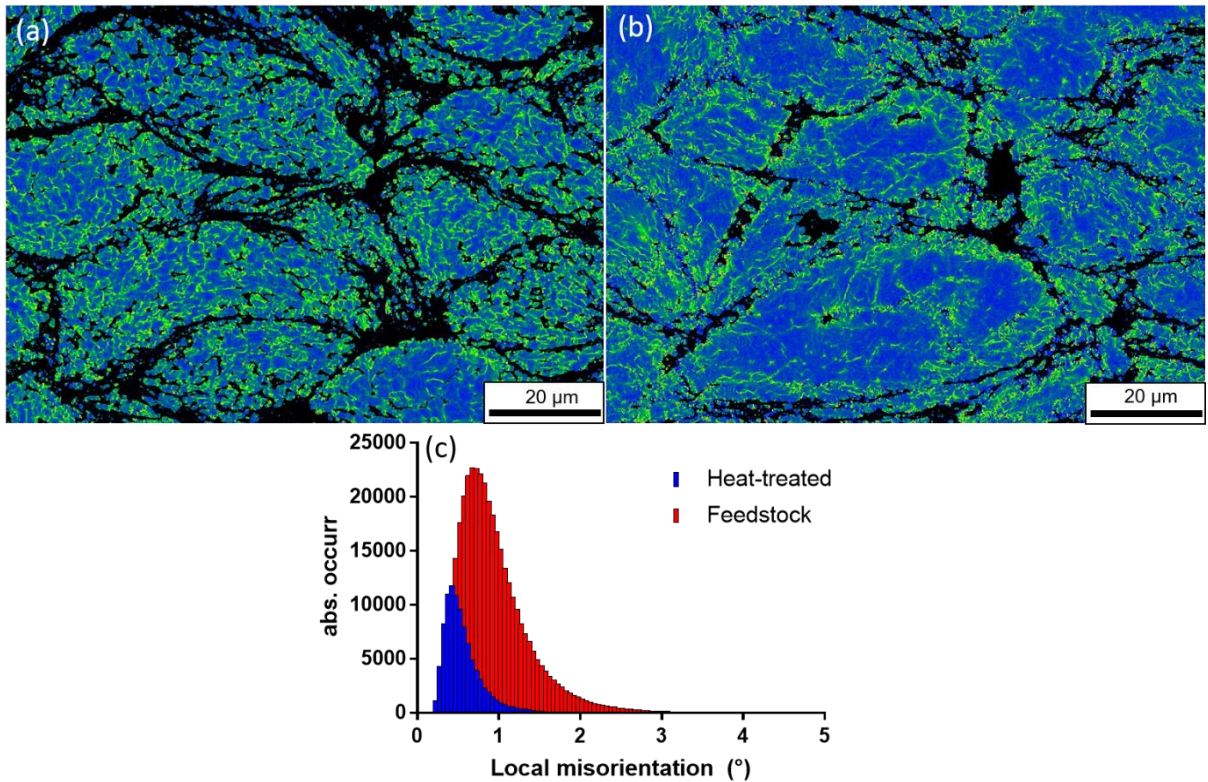

Fig. 13: Local average misorientation (LAM) maps of cold spray coatings deposited using (a) feedstock Al 6061 powder and (b) heat-treated powder, where the green and yellow regions represent the largest local misorientation. The feedstock particles exhibit a large misorientation in the particle interior whereas the solution heat treated particles display a localised misorientation at the particle-particle interface. The data represented (c) corresponds to the absolute number of local misorientation in both coatings.

The measurement of deformation and strain in a cold spray coating can provide valuable information. A quantitative analysis of this strain is not directly obtainable; however, Local average misorientation (LAM) maps, calculated using the average misorientation between each measurement and its 8 neighbours, can be plotted in order to get a visual representation of the strain in a material. It is represented as a gradient of colour, from blue showing an absence of lattice rotation to a green/yellow colour displaying a high degree of misorientation. The LAM is also used to measure the dislocation density, since misorientations are engendered by Geometrically Necessary Dislocations (GND), formed in a material to accommodate the plastic deformation and maintain a continuous lattice [39]. The LAM maps of the cold spray coatings deposited using the feedstock powder and the heat-treated powder are displayed Fig. 13. Overall, the misorientation appears to be more developed in the coating deposited using the feedstock powder (Fig. 13(a)), some areas can be observed with a more intense green colour, but the misorientation seems uniform in the whole coating. Quantitative data was plotted in Fig.13 (c) where the number of misorientation is represented for both coatings, confirming a major difference between the two deposits. However, when looking at the cold spray coating from the heat-treated powder (Fig. 13(b)), a clear trend emerges; the green colour representing the highest misorientation and thus the highest dislocation density is located at the extremities of the particles. A colour gradient is exhibited by all the splats— going from blue at their centre to a bright green at its extremities. This suggests a concentration of local misorientation and thus an accumulation of dislocations in those particle-particle interface areas, which the particle interior appears to be relatively free of.

**Discussion**

The homogenisation of powder microstructure through solution heat-treatment in recent studies [15–17] showed a major impact on cold spray deposition. Microstructural modification of gas atomised aluminium alloy powder by homogenisation resulted in a reduced porosity, higher deposition efficiency upon cold spraying, and a different dislocation and grain boundary size distribution. This section will cover the reasons leading to these behaviours based on the analysis of the microstructure of the powder and the coating. First, the microstructural modification of the powder particles will be discussed followed by a discussion on the bonding phenomena taking place in these two coatings.

The Al 6061 powder was homogenised by raising the temperature of the material above the solvus curve, as expected in a solution heat treatment process. The microstructural modification observed in previous studies on AA7075 [15] and AA2024 [16] was also noticed in the present study, accompanied by the formation of submicron precipitates after the heat-treatment. The dissolution of grain boundaries precipitates during solution heat treatment of gas atomised aluminium alloy powder 6061 was recently investigated on a chemical composition level. It was confirmed that the number of secondary phases was greatly reduced, especially the proportion of $Mg_2Si$, which drastically decreased after the first minutes of heat treatment [40]. Two different coatings microstructure resulted from the spraying of the feedstock and the heat-treated powder. As cold spray is based on the plastic deformation of the incident particles, porosity can be used as a deformation and bonding indicator. Junction voids and gaps between particles, are used to assess the optimum deposition conditions for a material. Often depending on spraying parameters, it can be corrected by increasing the gas temperature or pressure. In this study, the two coatings (Fig. 8) exhibit a major difference in porosity, although having been sprayed in the same conditions, which suggests an important modification in the particles' deposition behaviour.

The feedstock powder was found to contain a certain amount of LAGB (Fig. 7), located mostly towards the edge of the particle. LAGBs is often assumed to be composed of an array of dislocations, their presence in the powder is a potential indicator of residual stresses in the droplet post-atomisation [23,35,36]. In comparison, the solution heat-treated powder has a negligible amount of LAGB, coherent with the formation of the stress-free grains post heat-treatment. The reduction of LAGB in the heat-treated powder was attributed to the recrystallisation taking place during the heat-treatment. At the heat-treatment temperature the material recovers and recrystallises where new defects-free grains are formed. The equiaxed shape of the grain in the solution heat-treated powder is also an indicator of recrystallisation [43].

In this study, EBSD results (the band contrast (Fig. 9) and the local average misorientation (Fig. 13)) of the coating deposited from feedstock particle reveal that deformation and dislocation are distributed homogeneously across the whole splat. The large misorientation was observed in between each HAGB (Fig. 12) where various misorientation between 1 and 10° were detected in between different measured points. The presence of a wide distribution of LAGB in the centre of the splat is also an indication of heavy deformation in the splat interior, LAGB being often considered as an array of dislocations. On the other hand, LAGB in the coating deposited using the heat-treated powder appear to be distributed mainly at the splat interface. The local average misorientation data, typically used to measure the local strain variations, exhibits a gradient from blue (low misorientation) to green (high misorientation) from the centre of the heat-treated splat to its extremities. This is in contrast with the homogeneous strain distribution observed in the coating with the feedstock powder. The comparison of the misorientation graphs reveals a clear difference in the dislocation density after deposition.

During the high strain rate impact ($10^{6-9}$ s$^{-1}$) the localised temperature rise combined with the severe plastic deformation (SPD) results in phenomena such as adiabatic shear instability and dynamic recrystallisation which has been showed to play a significant role in particle bonding [18, 37]. The microstructure of the resulting coating is representative of the thermomechanical phenomena taking place during impact. Three different regions are commonly observed in a typical cold spray deposit. The first one, corresponding to the splat interiors, is relatively lightly deformed which can be easily observed in the present coating deposited using the solution heat-treated powder (Fig. 9) where the centre of the splat presents a bright colour indicating a high quality of Kikuchi bands and thus a relatively low deformation. The two following regions are located closer the particle-particle interface, where various mechanisms occur due to the extensive shear deformation. The most common model used to describe these phenomenon is called rotational dynamic recrystallisation [23]. In this model, the dislocations are propagated throughout upon impact and the lattice rotates along the shear direction, parallel to the particle-particle interface. The accumulation of dislocations in the interfacial area at temperature and strain exceeding certain values area then leads to the formation of elongated subgrains corresponding to the second region in a cold spray deposit microstructure. This phenomenon is subsequently followed by a subdivision of those subgrains into a third region. The rotation and movement of those subgrains to accommodate the severe plastic deformation during impact is at the origin of the ultrafine grains. This grain refinement in the interfacial area was observed in various alloys using TEM and also shown using EBSD and the grain boundary distribution in aluminium deposits [13], [21,39]. In that sense, the role of dislocations in the cold spray bonding mechanisms is critical, and their ability to accumulate and lead to this dynamic recrystallisation and the rotation of the subgrains is a pre-requisite for successful bonding. The presence of a clear dislocation accumulation in the particle-particle interfacial regions combined with the reduced porosity suggests that this phenomenon is responsible for a closer bond between the particles. In this study, the limited indexing in the interfacial area does not provide sufficient information to observe the recrystallisation and give an accurate measurement of the subgrain formation in this area. Nonetheless, the difference in dislocation density at the interfacial splat boundaries at the feedstock and the heat-treated coatings can be used at an indicator of the localised deformation in each particle.

A recent study showed that the presence of secondary sub-micrometric phases in metallic powder can act as nucleation sites for recrystallisation across the whole particle during cold spray deposition [24]. The accumulation of dislocations around the secondary phases led to dynamic recrystallisation inside the particle, which reduced the number of dislocations in the particle- particle interfacial area due to a more homogeneous deformation throughout the particle. This phenomenon of localised dislocation accumulation has been also observed in cases of severe plastic deformation of various materials during processes such as equal channel angular pressing [46] or accumulative roll bonding [47]. The interaction of dislocation with precipitates and second-phase particles was thoroughly studied using TEM, where a high dislocation density was observed in their surrounding area [48]. Dynamic recrystallisation and grain refinement was thus observed in the vicinity of the particles due to the dislocation accumulation in these regions. In this study, the presence of an intermetallic network along with a large number of precipitates in the feedstock powder are believed to act in a similar way, hindering the dislocation movement during deposition.

The solution heat-treated powder, appearing free of dislocations/ lattice defects and exhibiting a larger grain structure compared to the feedstock powder, showed an improvement in porosity once deposited (Fig. 8). The decreased proportion of precipitates that can be observed in the particles cross-section (Fig.3) was recently quantified as a function of heat treatment time [40]. It was confirmed that a thermal processing was reducing the quantity of the majority of the precipitates that can be found in as-atomised particles. The particles, presenting a much lower density of precipitates, are likely to facilitate the accumulation of these dislocations in the interfacial area as evidenced by Fig. 13. During the severe plastic deformation occurring during deposition, this localised deformation at the particle-particle boundaries enhances the formation of elongated grains and further subgrains formation takes place. As described in the rotational dynamic recrystallisation model, the rotation of those subgrains to accommodate the plastic deformation is crucial during deposition. The proposed mechanism occurring during deposition of both feedstock and heat-treated powder is presented in the schematic Fig.14. The particles are impacting on a surface presented as flat, which does not correspond to the realistic conditions of a cold spray deposition. This schematic compares and explains the behaviour of two different particles impacting in the same conditions.

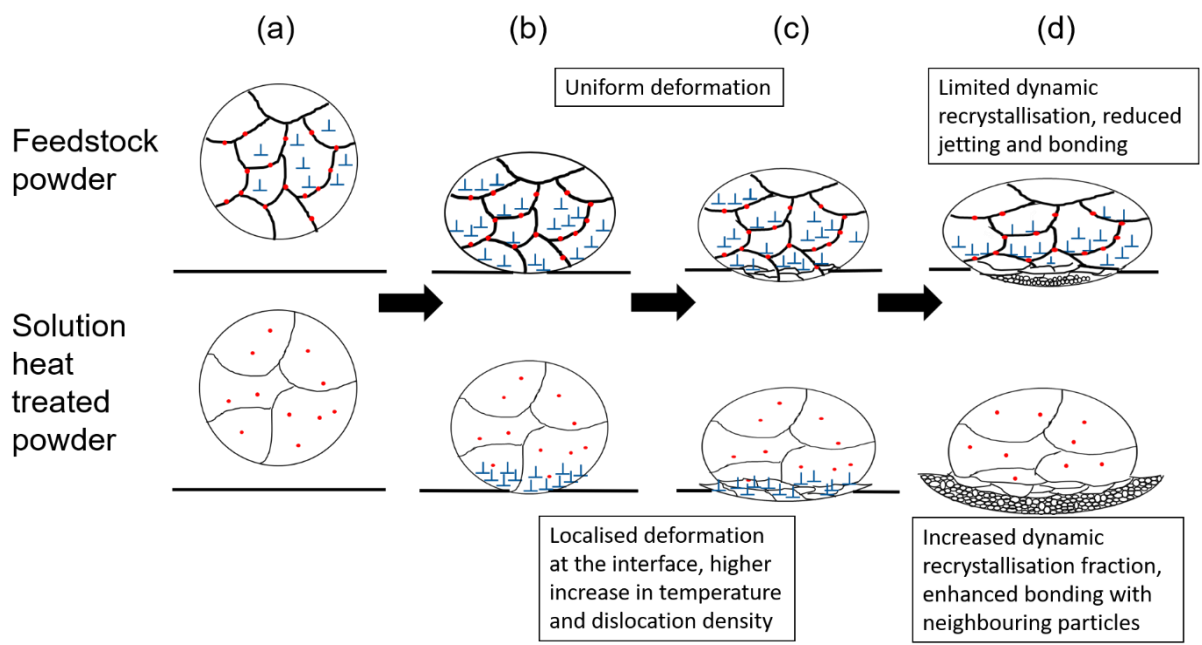

Fig. 14: Schematic evolution of a feedstock and a solution heat-treated particle impacting during cold spray deposition. The interdendritic region is represented as thick black lines whereas the thinner black lines represent the grain boundaries in both particles—the interdendritic region is superposed with the grain boundaries in the first case. Dislocations are represented in blue, and precipitates are represented as red dots in both powders. The intermetallic network composed of numerous precipitates in the feedstock particle leads to an accumulation of the dislocations in the particle interior as well as at the particle-particle interface. This limits the dislocation density at the particle-particle interface thus the dynamic recrystallisation and subgrain rotation. On the contrary, the schematic deformation of the solution heat-treated particleat the point of contact shows a localised deformation at the interface and an accumulation of dislocations. This accumulation of dislocations leads to subgrain formation and rotation during impact to accommodate plastic deformation and enhance the bonding between adjacent particles.

If the presented model is correct, as soon as the particle hits the substrate, a high density of dislocations is generated by the high impact pressure in the interfacial area (Fig.14(b)). In the case of the feedstock powder, the plastic deformation is accommodated homogeneously throughout the particle due to the presence of a thick intermetallic network and precipitates in the particle microstructure, as well as the presence of original dislocations in the material. The solution heat treated particle, however, constituted of a rather homogeneous matrix accompanied with s precipitates, is deformed locally in the interfacial area, at the origin of the impact. The enhanced formability of the particles due to the dissolution of the interdendritic structure led to a more intense deformation in the contact zone rather than a homogeneous flattening of the splats. The free moves of dislocation across the incident heat-treated particles gives the possibility of this localised deformation". The accumulation of dislocations in the bottom-part of the particle Fig.14 would be the origin of the grain elongation in the interfacial region (Fig.14(c)). The high number of dislocations in this localised area enhanced the formation of elongated grains and the subsequent formation of subgrain which are further rotated to accommodate the plastic deformation. It is believed that this mechanism was made possible by the temperature increase at the interfacial area upon impact. In cold spray, the plastic deformation is such that temperature raises drastically in a localised area. The temporal evolution of a cold sprayed Al particle of 35 µm diameter during impact was modelled by Bae et al., and a sudden temperature rise of 600 K to 1100 K was showed depending on the area and the spraying parameters as a result of the severe plastic deformation occurring and the followed adiabatic shear instability leading to the material flow at the interface [18]. In contrast, the limited dislocation accumulation of the feedstock particle provided a limited opportunity for subgrain rotation.

**Conclusions**

The present study details the modification of a gas atomised Al 6061 powder by solution heat-treatment and its effect on subsequent cold spray deposition. SEM and EBSD techniques were used to investigate the microstructural evolution occurring during deposition. The following conclusions can be drawn from this study:

- The feedstock powder underwent a recrystallisation process during the heat-treatment, leading to the formation of larger equiaxed grains post-process. An absence of LAGB prior to spraying was also observed in the heat-treated particles.
- A more intimate bonding was observed between the particles in the coating deposited using the solution heat treated powder, whereas gaps and inter-particles voids were characterised in the coating sprayed using the feedstock powder.
- The coating deposited from feedstock powder exhibited a homogeneous distribution of dislocations and misorientations across the whole deposit as revealed by the misorientation lines and the local average misorientation graph. In contrast, the coating deposited from solution heat treated powder appeared to be relatively free of dislocations in the particle interiors whereas the particle extremities exhibited a large number of misorientation.
- The dislocation distribution was found to be correlated with the particle-particle bonding in the deposit. An accumulation of dislocations in the particle-particle interfacial area enhanced the dynamic recrystallisation in the solution heat-treated powder; on the other hand, the dislocation movement hindered by the grain boundaries and the intermetallic network acting as pinning points in the feedstock powder limited their accumulation in the particle-particle interfacial area and thus limited the accommodation of plastic deformation.

This study provides an understanding of the benefits of the heat-treatment of aluminium alloy powder prior to cold spray. It gives the possibility of considering an extended use of this process in industry using larger scale facilities as an improvement of the resulting coatings properties for similar process parameters is highly beneficial to any application.

**Data availability**

All the data used for this study have been included in the document.

**Acknowledgments**

This work was supported by the Engineering and Physical Sciences Research Council [grant number EP/M50810X/1]; in the form of a CASE PhD studentship and industrial funding from TWI via the National Structural Integrity Research Foundation (NSIRC). The authors also acknowledge support from Ms Heidi De Villiers Lovelock and Dr Henry Begg at TWI for editorial work and valuable technical discussion, Mr Rory Screaton at the University of Nottingham for conducting the cold spray experiments, and Mr Kamaal Al-Hamdani for providing samples for experiments.